\documentclass[aps,prl,reprint,superscriptaddress,twocolumns]{revtex4-1}
\usepackage{amsmath}
\usepackage{graphicx}
\usepackage{bm}
\usepackage{dcolumn}

\begin{document}

\title{Containing intense laser light in circular cavity with magnetic trap door}


\author{X. H. Yang}\thanks{Electronic mail: xhyang@nudt.edu.cn}
\affiliation{College of Science, National University of Defense Technology, Changsha 410073, China}

\author{W. Yu}
\affiliation{Shanghai Institute of Optics and Fine Mechanics, Chinese Academy of Sciences, Shanghai 201800, China}

\author{M. Y. Yu}\thanks{Electronic mail: myyu@zju.edu.cn}
\affiliation{Institute for Fusion Theory and Simulation, Zhejiang University, Hangzhou 310027, China} \affiliation{Institut f\"{u}r Theoretische Physik I, Ruhr-Universit\"{a}t Bochum, D-44780 Bochum, Germany}

\author{H. Xu}\affiliation{School of Computer Science, National University of Defense Technology, Changsha 410073, China}

\author{Y. Y. Ma}
\affiliation{College of Science, National University of Defense Technology, Changsha 410073, China}

\author{Z. M. Sheng}
\affiliation{Key Laboratory for Laser Plasmas (Ministry of Education), Department of Physics and Astronomy, Shanghai Jiao Tong University, Shanghai 200240, China}\affiliation{SUPA, Department of Physics, University of Strathclyde, Glasgow G4 0NG, UK}

\author{H. B. Zhuo}\affiliation{College of Science, National University of Defense Technology, Changsha 410073, China}

\author{Z. Y. Ge}\affiliation{College of Science, National University of Defense Technology, Changsha 410073, China}



\author{F. Q. Shao}\affiliation{College of Science, National University of Defense Technology, Changsha 410073, China}


\date{\today}

\begin{abstract}
It is shown by particle-in-cell simulation that intense circularly
polarized (CP) laser light can be contained in the cavity of a
solid-density circular Al-plasma shell for hundreds of light-wave
periods before it is dissipated by laser-plasma interaction. A
right-hand CP laser pulse can propagate almost without reflection
into the cavity through a highly magnetized overdense H-plasma slab
filling the entrance hole. The entrapped laser light is then
multiply reflected at the inner surfaces of the slab and shell
plasmas, gradually losing energy to the latter. Compared to that of
the incident laser, the frequency is only slightly broadened and the
wave vector slightly modified by appearance of weak nearly isotropic
and homogeneous higher harmonics.
\end{abstract}

\pacs{42.65.Jx, 52.25.Xz, 52.65.Rr}

\maketitle

Recently, intense laser interaction with plasma has been widely
studied in connection with applications such as inertial confinement
fusion \cite{Lindl14} and laser-driven particle acceleration
\cite{Macchi13,Yang14,Yang15P}. Particle-in-cell (PIC) simulations
and theoretical studies \cite{Naumova01,Yu78} have shown that
isolated intense light pulses on the scale of the laser wavelength
can be self-consistently trapped in plasmas as solitons. Trapping of
electromagnetic (EM) waves in plasma can also be from excitation of
surface plasma waves \cite{Stenflo96,Bakunov97,Kumar07}, beam-plasma
and/or parametric instabilities \cite{Kaw73,Silin77}. Recently, it
is found that an intense laser pulse that has entered into a hollow
shell through a hole can also be trapped when a near-critical
density plasma layer is applied or self-generated to seal the
entrance hole \cite{Luan15}. In all these cases, there is much loss
of the laser energy to the plasma. On the other hand, there has also
been considerable research on trapping and localizing light in
atomic media or photonic structures
\cite{Storzer06,Hau99,Fleischhauer00,Tanabe07,Melentiev13}. However
in these works no intense laser-plasma interaction is involved
because of the relatively weak laser intensities used.

Intense magnetic fields of $10^3-10^5$T and above are ubiquitous in
the cosmic environment \cite{Canuto71}. Strong magnetic fields can
also be self-generated during ultraintense-laser interaction with
matter by compression of seeded magnetic fields
\cite{Knauer10,Chang11,Yoneda12}, baroclinic effect \cite{Wagner04},
high-current electron beams \cite{Fujioka13,Korneev15}, etc. In the
presence of kilotesla magnetic fields, the electron gyrofrequency
$\omega_c$ can be comparable to or even larger than the laser
frequency $\omega_L$. In this case, a right-hand circularly
polarized (RHCP) laser can propagate deep into an overdense plasma
along an intense embedded magnetic field without encountering cutoff
or resonance \cite{Chen06,Yang15}. Moreover, transmission of RHCP
laser into dense plasma can be controlled by an intense magnetic
pulse in the plasma \cite{Ma16}.

In this Letter, we propose a scheme for trapping intense light in a
hollow shell of high density plasma by sending a RHCP laser light
pulse through a strongly magnetized slab that serves as a
transparent trap door. The laser pulse can propagate through the
slab with almost no reflection. Once the entire laser pulse (which
can be suitably long) is inside the shell cavity, the magnetic field
in the slab is turned off, so that the light becomes contained. It
is then multiply reflected by the unmagnetized shell plasma or
propagates along its surface. As a result, the light can survive for
hundreds of light-wave periods until all its energy is absorbed by
the slab and shell plasma.

The optical properties of a linear medium can be described by its
refractive index $N=ck_L/\omega_L=\varepsilon^{1/2}$ \cite{Chen06},
where $k_L=2\pi/\lambda_L$, $\omega_L$, and $c$ are the wave number,
frequency, and vacuum speed of the light and $\varepsilon$ is the
dielectric constant of the medium. For simplicity, in the following
we shall normalize the plasma density $n_e$ by the critical density
$n_c=m_e\omega_L^2/4\pi e^2$, where $e$ and $m_e$ are the electron
charge and mass, respectively. Accordingly, for unmagnetized plasma
we have $\varepsilon=1-\omega_p^2/\omega_L^2=1-n_e$. For magnetized
plasma with the external magnetic field $\bm{B}_0$ (normalized by
$m_ec\omega_L/e$) along the laser propagation direction, the
dielectric constant can be written as $\varepsilon_B=1-n_e/(1\pm
B_0)$ \cite{Yang15,Chen06}, where the plus and minus signs
correspond to the so-called $L$ and $R$ electromagnetic waves,
respectively \cite{Chen06}. Thus, for the $R$ wave $\varepsilon_B$
is always larger than unity if $B_0>1$, and the plasma is
transparent at any density. In contrast, the $L$ wave has a cutoff
density at $n_L=1+B_0$, where it will be reflected.

\begin{figure}\suppressfloats\centering
\includegraphics[width=8.5cm]{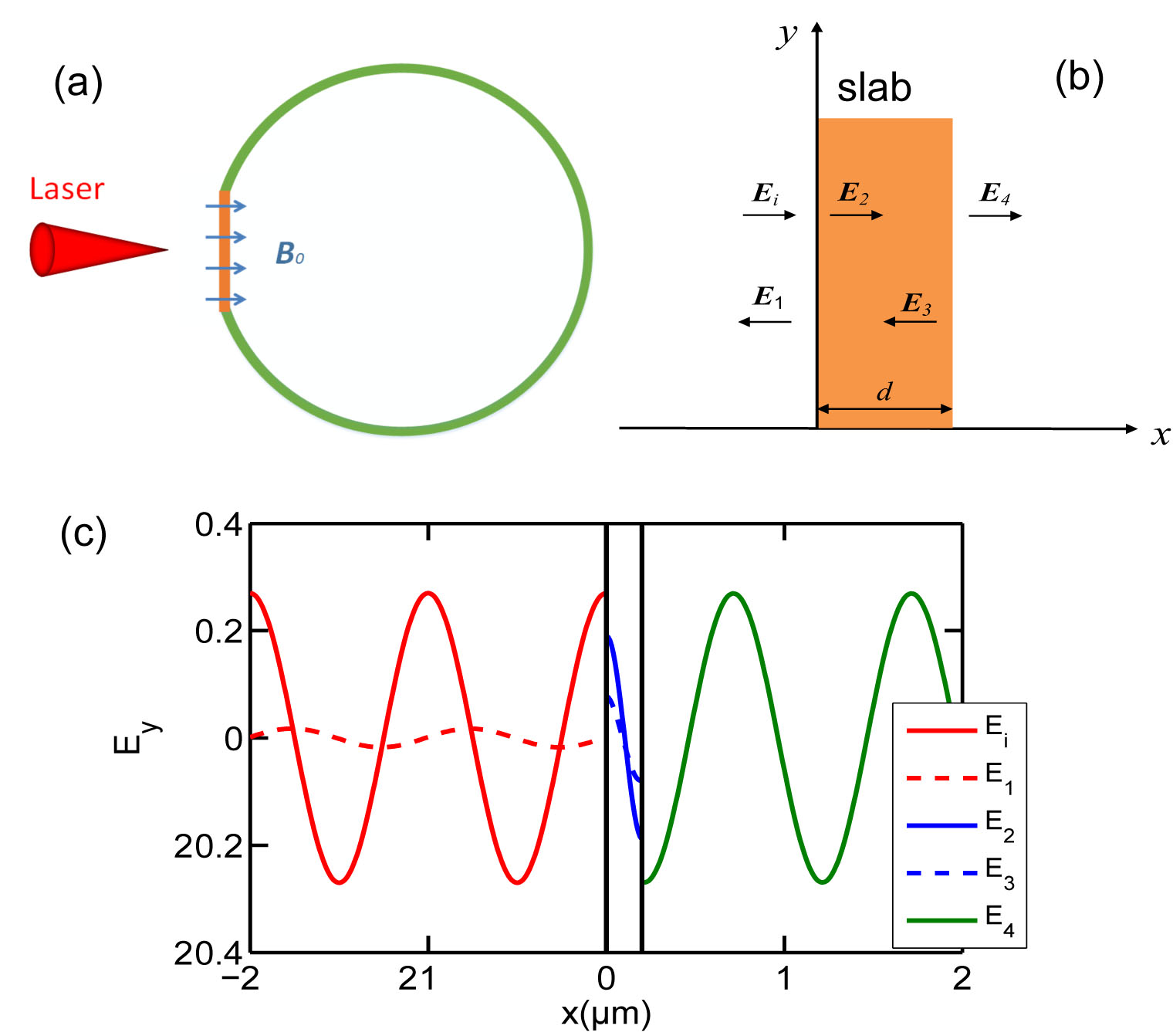}%
\caption{\label{f1} Schematic of (a) the proposed scheme. The
entrance hole on the left shell wall is fitted with an overdense
flat slab, which is strongly magnetized until the laser pulse has
entered the cavity, and (b) the one-dimensional model for the slab
used for illustrating the $R$ wave propagation. (c) Analytical
solution for linear propagation of the $R$ wave through the slab.
Here and in the following figures, $E_y$ is normalized by
$m_ec\omega_L/e$ (corresponding to $3.22\times10^{12}$V/m).}
\end{figure}

The proposed trapping scheme is illustrated in Fig. \ref{f1}(a). The
circular plasma shell has on its left wall a small section replaced
by a highly-magnetized overdense flat slab that serves as a trap
door for the incident laser pulse. In order to see the function of
the slab, it is instructive to first consider the one-dimensional
linear theory of light transmission through a thin highly-magnetized
high-density plasma layer, as shown in Fig. \ref{f1}(b). The total
wave electric field $E_y(x)$ in each of the three regions can be
written as \cite{Kong90,Scharstein92}
$E_i\exp{(-ik_Lx)}+E_1\exp{(ik_Lx)}$ for $x<0$,
$E_2\exp{(-ik_sx)}+E_3\exp{(ik_sx)}$ for $0<x<d$, and
$E_4\exp{[-ik_L(x-d)]}$ for $x>d$, where $E_i$ and $E_1$, $E_2$,
$E_3$, and $E_4$ are the amplitudes of the laser light incident on
and reflected from the front surface of the slab, transmitted into
the slab, reflected from the back surface of the slab, and
transmitted into the vacuum, respectively,
$k_s=\varepsilon_B^{1/2}k_L$ is the wave number in the slab.
Applying the boundary conditions that at each of the boundaries (the
total) $E_y$ and $\partial_xE_y$ should be continuous, one obtains
after some algebra
\begin{equation} \label{field3}
\begin{split}
   E_1=&\frac{-i2RE_0\sin(k_sd)}{R^2e^{ik_sd}-e^{-ik_sd}},\\
   E_2=&\frac{2RE_0}{R(N_s+1)-(N_s-1)e^{-i2k_sd}},\\
   E_3=&\frac{2E_0}{R(N_s+1)e^{i2k_sd}-(N_s-1)},\\
   E_4=&\frac{4N_sE_0}{(N_s+1)^2e^{ik_sd}-(N_s-1)^2e^{-ik_sd}}.
\end{split}
\end{equation}
where $N_s=\varepsilon_B^{1/2}$ and $R=(N_s+1)/(N_s-1)$. Additional
internal reflections are negligible. Fig. \ref{f1}(c) shows the
profile of the electric field obtained from Eq. (1) for $n_e=20n_c$,
$d=0.2\mu$m, $B_0=5$, and $E_i=0.27$ (the same as the corresponding
parameters used in the PIC simulation below). It can be seen that
the reflected light ($E_1=0.017$) is only 6\% of that ($E_i=0.27$)
of the incident, and the transmitted light ($E_4=0.269$) is very
close to the latter.

The relativistic 2D3V PIC simulation code EPOCH \cite{Arber15} is
used to investigate our scheme. The shell consists of Al plasma (ion
mass $m_{Al}=26.98m_p$, where $m_p=1836m_e$ is the proton mass) with
charge $Ze=10e$ and density $50n_c$. Its inner radius and thickness
are $14.8\mu$m and $0.2\mu$m, respectively. The slab consists of
hydrogen plasma with density $20n_c$. Its height and thickness are
$10\mu$m and $0.2\mu$m, respectively. The center of the shell cavity
is located at $(x,y)=(20,20)\mu$m, so that the left front of the
slab is at $x=5.86\mu$m. An external magnetic field of strength
$B_0=5$ (corresponding $5.36\times10^4$T in dimensional units) along
the laser axis is embedded in the slab plasma, which is shut down
after the laser pulse injection is complete. To account for
laser-plasma interaction induced plasma expansion, the extent of
$\bm{B}_0$ is slightly wider than the slab. The initial temperature
of the slab and shell is 100 eV. The simulation box is
$40\mu$m$\times40\mu$m with $4000\times4000$ cells. Each cell
contains 50 macro particles per species. At $t=0$, a RHCP laser
pulse of wavelength $\lambda=1\mu$m and intensity
$I_0=2\times10^{17}$W/cm$^2$ (or laser parameter
$a_L=eE_L/m_ec\omega_L=0.27$, where $E_L$ is the peak laser electric
field) enters from the left side ($x=0$) of the simulation box and
is focused on of the slab at $x=5.86\mu$m. Its front rises with a
Gaussian profile for 17 fs to the peak intensity and remains
constant for 150 fs. The transverse profile of the laser pulse is
Gaussian, with spot radius $2\mu$m.

\begin{figure}[h]\suppressfloats\centering
\includegraphics[width=8.5cm]{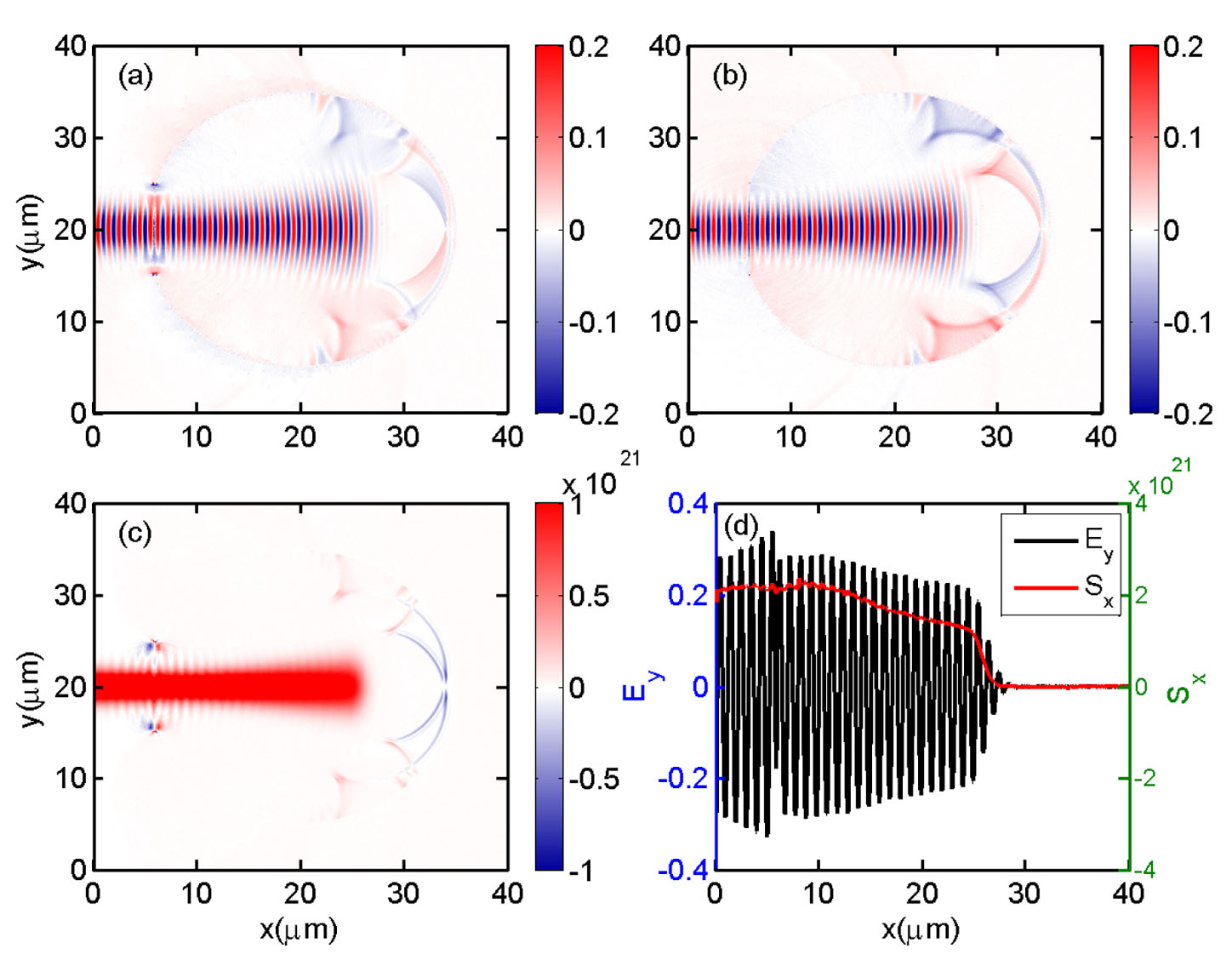}%
\caption{\label{f2} Distribution of the transverse (a) electric
($E_y$) and (b) magnetic ($B_z$) fields, and (c) the axial Poynting
flux $S_x$ at $t=100$fs. (d) $E_y$ and $S_x$ (averaged over
0.5$\mu$m around $y=20\mu$m) along the $x$ direction. Here and in
the following figures, $B_z$ is normalized by $m_ec\omega_L/e$
(corresponding to $1.07\times10^{4}$T) and $S_x$ is in units of
W/m$^2$. The highly localized $E_y$ and $S_x$ fields at the top and
bottom slab-shell boundaries can be attributed to the complex
laser-plasma interaction there. See also Figs. \ref{f3}(c) and
\ref{f4}.}
\end{figure}

Figure \ref{f2} shows the electric field $E_y$, magnetic field
$B_z$, and axial Poynting vector $S_x$ at $t=100$fs. We see that the
laser pulse can easily enter the overdense slab into the cavity. It
propagates through the slab almost without reflection, and the
amplitude of the transmitted laser in the vacuum is very close to
that of the incident laser, in good agreement with the theoretical
analysis (see Fig. \ref{f1}(c)). One can see from Figs. \ref{f2} for
$t=100$fs that there is little energy loss by the laser as it passed
the slab. The axial Poynting flux is everywhere positive except at
the shell-slab boundary, where some light is scattered.

\begin{figure}\suppressfloats\centering
\includegraphics[width=8.5cm]{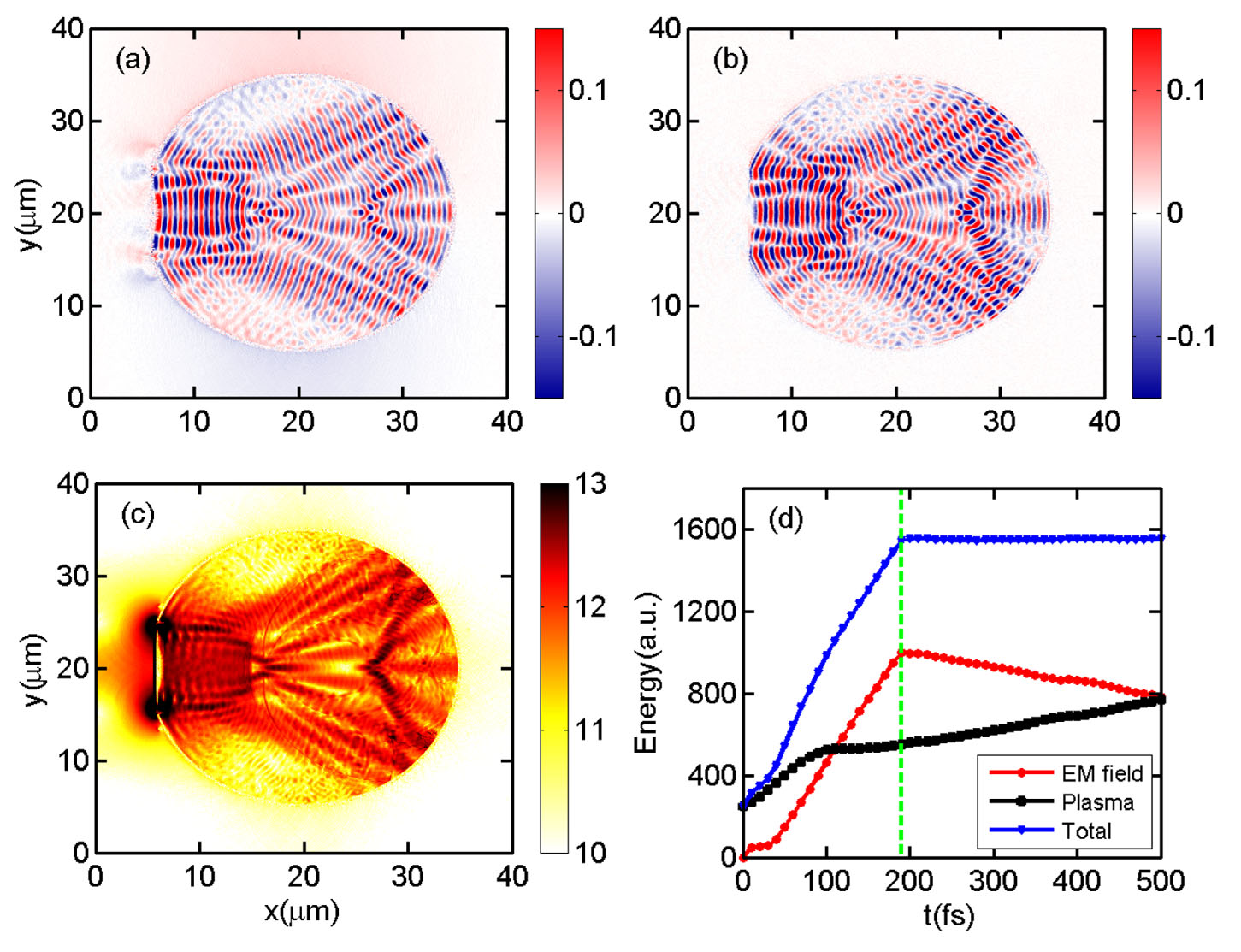}%
\caption{\label{f3} Distribution of (a) $E_y$, (b) $B_z$, and (c) EM
field energy density (in J/m$^3$) at $t=350$fs. (d) Evolution of the
energies of the light waves in the shell (red circles), the plasma
(including the shell and slab, black squares) and the sum of the
former energies (blue triangles). The green dashed line in (d) marks
the termination of the laser pulse injecting into the shell. Recall
that the latter has a short ($17$fs) Gaussian front, followed by a
long ($150$fs) flat-top tail.}

\end{figure}

Reflections from the cavity wall and wave interference lead to
restructuring of the EM fields in the cavity as well as loss of wave
energy to the slab and shell plasmas. Figure \ref{f3} for $t=350$fs
shows the transverse electric field $E_y$, magnetic field $B_z$, and
EM field energy density
$\mathcal{W}=\frac{1}{2}(\epsilon_0|\bm{E}|^2+{\mu_0^{-1}}|\bm{B}|^2)$,
where $\epsilon_0$ and $\mu_0$ are the free-space permittivity and
permeability, respectively. We see that light is trapped in the
cavity and its structure is rather complex. Because of constructive
interference of the reflecting light in the cavity, locally the
magnitudes of the wave electric and magnetic fields can be up to
1.61 and 2.35, respectively, times that of the incident laser.
Figure \ref{f3}(d) for the evolution of the energies of the trapped
light and the plasmas shows that the former increases with time
until the laser pulse has completely entered the cavity (at
$t\sim190$fs, recall that the slab is $5.86\mu$m away from the left
side of the simulation box). Then it gradually decreases. One can
also see that the energy of the plasmas first increases rapidly,
mainly due to light absorption by the slab electrons, which is
heated and expands, as shall be discussed later. Figure \ref{f3}(d)
also shows that after the laser pulse has completely entered the
cavity at $t\sim190$fs, energy absorption by the plasma continues at
a slower rate, and the total energy of the light in the cavity and
the plasma in the slab and shell is well conserved.

On the other hand, Fig. \ref{f3}(d) also shows that even though
transfer of light energy to the slab and shell plasmas starts as
soon as the laser enters the slab, even at $t=500$fs the trapped
light still retains about $78.6\%$ of its maximum energy, namely that
at $t\sim190$ fs. We have not attempted to optimize the light
trapping, which has to be done by trial and error adjustment of the
initial parameters. For example, simulations indicate that the
energy of the trapped light can increase with the cavity size. It
should also be possible to improve the design of the slab so that
absorption of reflected light by its cavity facing side is
minimized.

\begin{figure}\suppressfloats\centering
\includegraphics[width=5.5cm]{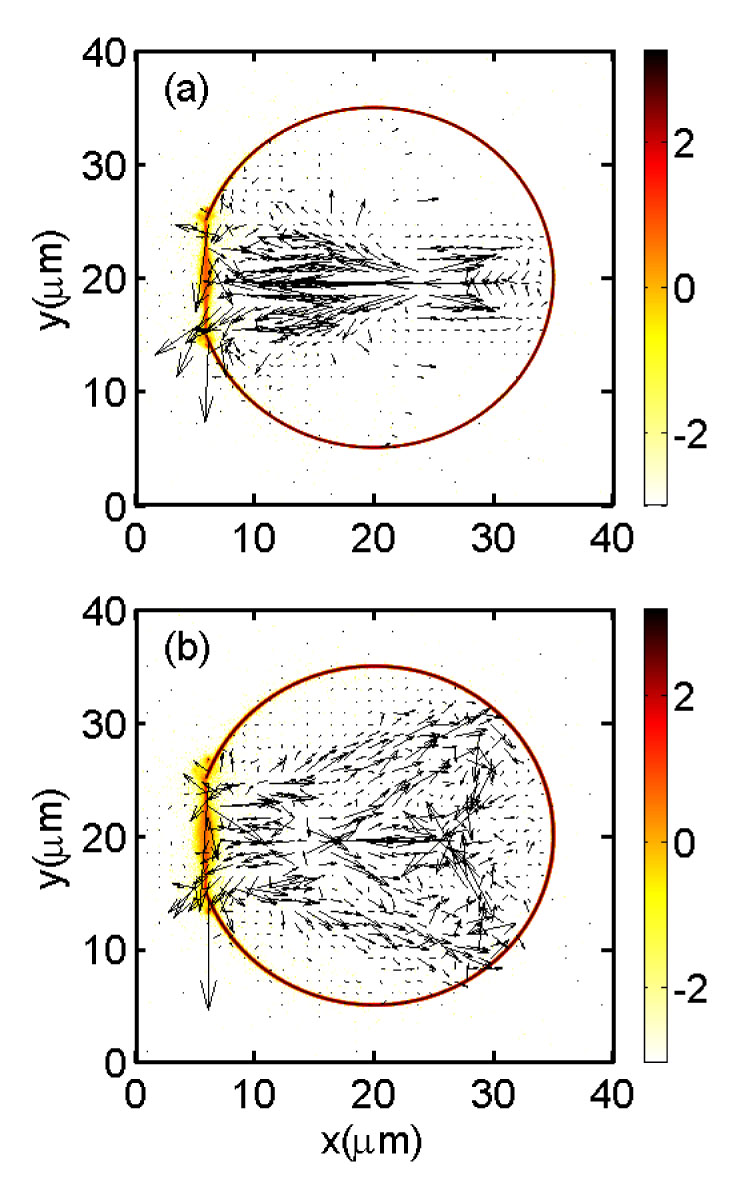}%
\caption{\label{f4} Distribution of the instantaneous Poynting flux
$\bm{S}=\bm{S}_x+\bm{S}_y$ (arrows, their lengths indicating the
magnitudes), and the electron density (color coded) of the slab and
the shell at (a) $t=250$fs and (b) $t=350$fs.}
\end{figure}

Figure \ref{f4} for the plasma density and light Poynting flux
distributions shows that the slab-shell boundaries are modified and
strong local electric and magnetic fields are generated around them.
One can also see in Fig. \ref{f4}(a) for $t=250$fs that, as
expected, at this instant the flow of light energy is still mainly
around the $\pm x$ directions. Figure \ref{f4}(b) for $t=350$fs shows
that at this later time interaction of the trapped light with plasma
takes places mainly in three regions: at the slab and two regions on
the shell wall. That is, the Poynting flux is at first localized
along the axis region, but then spreads to other regions at latter
times. Focusing and defocusing of the reflected light on the cavity
axis can be observed at $x\sim 28\mu$m and $x\sim 17\mu$m. The $20
n_c$ highly magnetized H slab plasma at the laser entrance is
strongly heated and expands, but the very high charge density Al
shell plasma remains almost unaffected. The reason is that for the
level of light intensity here, energy absorption by the shell plasma
is mainly due to vacuum heating \cite{Brunel87,Gibbon96}, whose
efficiency depends on the skin depth, which for the $50n_c$ $Z=10$
Al shell plasma is 5 times smaller than that of the $20n_c$ H slab
plasma, where efficient resonant absorption dominates, especially
after the expansion \cite{Gibbon96,Friedberg72}. At still later
times the interaction regions on the right inner shell wall are
wider spread, but the structure and behavior of the trapped light in
the cavity remain similar. Due to continuous energy transfer to the
slab and shell plasma, after the laser pulse has fully entered the
cavity at $t\sim190$ fs, the total light/plasma energy
decreases/increases almost linearly with time, but as mentioned, the
total energy remains constant. However, no stationary or
quasistationary state was found even at $t\gtrsim500$ fs, when the
light field has become weak.

\begin{figure}[h]\suppressfloats\centering
\includegraphics[width=8.7cm]{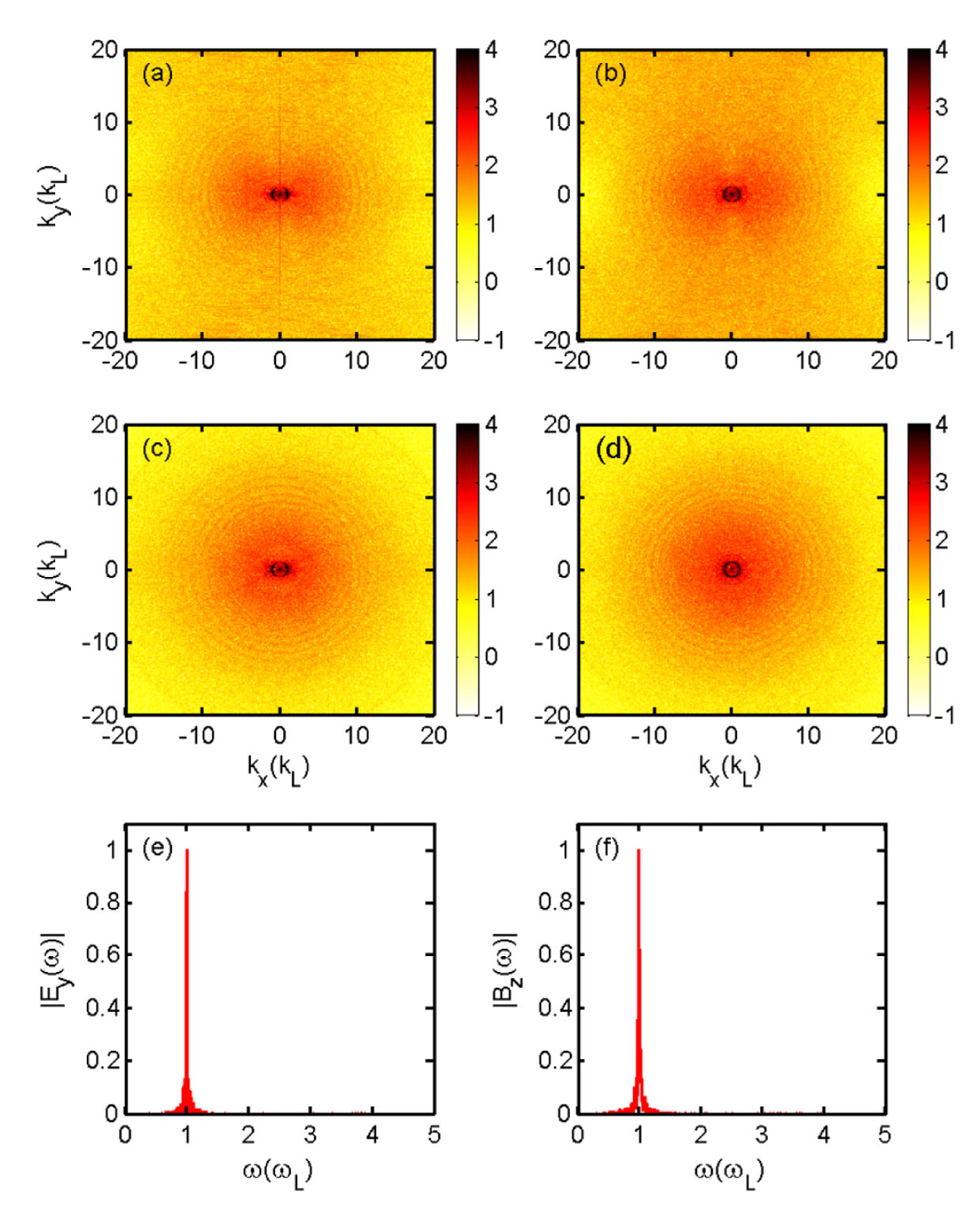}%
\caption{\label{f5} Fast Fourier transform of $E_y$ (top row) and
$B_z$ (bottom row) at $t=300$ fs (left column) and 400 fs (right
column). Frequency spectra $|E_y(\omega)|$ (e) and $|B_z(\omega)|$
(f) (arbitrary units) at the center ($x=20\mu$m, $y=20\mu$m) of the cavity.}
\end{figure}

Figure \ref{f5} shows the spatial EM field spectra $E_y(\bm{k})$ and
$B_z(\bm{k})$ at two instants after the laser pulse has fully
entered the cavity. One can see that even though the spatial wave
structure in the real space is rather complex, its Fourier space is
relatively simple. As mentioned, initially the $\pm k_x$ components
from the just-transmitted laser light dominates (not shown), but
$k_y$ components also appear as reflections take place since the
laser is of finite spotsize and the cavity wall is curved. Figures
\ref{f5}(a)--(c) show that except around $\theta\sim\pi/2$ and
$3\pi/2$ (apparently due to less light reflections occurring around
these angles in the physical space), the $|\bm{k}|=1$ mode (small
dark partial circles in the figures) associated with that
($\bm{k}=1\hat{e}_x$) of the input laser remains dominant at all
times, which can also be roughly seen from the angles of the
Poynting vectors in Fig. \ref{f4}. In fact, close examination shows
that the $|\bm{k}|<1$ region is more highly populated around the
$\pm x$ directions, as to be expected. Weak but distinguishable
spatial harmonics $|\bm{k}|=2, 3, 4, ...$ can also be observed. The
spectra continue to evolve with time as energy loss to the slab and
shell plasma continues, but their overall profiles remain almost
unchanged. In particular, they remain peaked at $|\bm{k}|=1$ (except
around $\theta\sim\pi/2$ and $3\pi/2$) and $k=0$. Figures
\ref{f5}(e) and (f) show the frequency spectra $E_y(\omega)$ and
$B_z(\omega)$ at the center of the cavity. At other locations in the
cavity they (not shown) are very similar. We see that the frequency
spectrum is highly peaked at the incident laser frequency
$\omega_L$, which together with the the dominance of the wavelength
near that of the incident laser, confirms that the light of the
latter is trapped without altering its basic characteristics. The
slight frequency broadening can be attributed to the presence in the
cavity of low-density electrons driven out by the laser-wall plasma
interaction, so that the vacuum light-wave dispersion relation is
slightly modified by inclusion of the plasma frequency $\omega_p$
($\ll k_Lc$). However, as already pointed out, we were unable to
identify the EM field or Poynting vector structures with that of an
eigenmode solution of the vacuum wave equation in circular geometry
\cite{Padgett95}, even though the trapped light has quite well
defined frequency and wavelength, as well as $E_y(\omega,\bm{k})$
and $B_z(\omega,\bm{k})$ relationship. The reason could be that
there is insufficient time to form an eigenmode, the cavity is not
really circularly symmetric because of the slab, and/or the
wave-plasma interaction at the slab and cavity wall cannot be
ignored. Furthermore, one can see in Fig. \ref{f5}(a)-(d) that the
region near $k=0$ is also enhanced, which can be attributed to the
multiple light reflections at the slab and shell plasma boundaries,
where $\bm{k}$ changes sign. Finally, we note that weak oscillations
(not shown) are also detected in the electron density, which can be
attributed to excitation of standing electron plasma oscillations by
the laser-plasma interaction.

In conclusion, we have considered a scheme for containing intense
light in the cavity of a solid-density Al plasma shell. PIC
simulations demonstrate that the laser light can enter the cavity
with the help of an axial magnetic field embedded in an overdense H
slab covering a hole in the shell. Inside the cavity the light wave
is reflected as well as partially absorbed by the slab and shell
plasma. Because of the multiple reflections, the wave frequency is
slightly broadened but remains close to that of the incident laser.
The wave vector, originally only along the laser axis, also acquires
new directions and weak spatial harmonics are generated, which then
become azimuthally isotropic except around $\theta\sim\pi/2$ and
$3\pi/2$. However, the wavelength of the incident laser still
remains the dominating spatial scale. Because of energy transfer to
the slab and shell, the trapped light continues to decay until it
vanishes. The results here may be useful for understanding storage
of intense light, as well as interpretation of phenomenon associated
with highly localized light \cite{Stenhoff99,Wu16}.

\begin{acknowledgments}\suppressfloats
We would like thank S. H. Luan for useful discussions. This work was
supported by the NNSFC (11305264, 11275269, 11374262, 11375265, and
91230205) and the Research Program of NUDT.

\end{acknowledgments}

\end{document}